\documentstyle[multicol,aps]{revtex}
\begin{document}
\title{Dynamics of Electrons in Graded Semiconductors}
\author{Michael R. Geller}
\address{Department of Physics, Simon Fraser University, Burnaby B.C.  
V5A 1S6, Canada}
\date{\today}
\maketitle

\begin{abstract}
I present a theory of electron dynamics in semiconductors with slowly
varying composition.
I show that the frequency-dependent conductivity, required for
the description of transport and optical properties, can be obtained
from a knowledge of the band structures and momentum matrix elements
of {\it homogeneous} semiconductor alloys. 
New sum rules for the electronic oscillator strengths,
which apply within a given energy band or between any two bands,
are derived, and a general expression for the width of the
intraband absorption peak is given. Finally, the low-frequency
dynamics is discussed, and a correspondence with the semiclassical
motion is established.
\end{abstract}

\pacs{PACS: 73.40.Hm, 71.27.+a}
\begin{multicols}{2}

Bloch's seminal analysis of the quantum 
mechanical motion of a particle in a periodic potential
lies at the heart of the modern electron theory of solids, 
and provides a basis---band theory---for the
entire conceptual framework of the subject. 
The present work is motivated by the tremendous interest
in the electronic properties of nanometer-scale 
semiconductor structures \cite{Bastard etal},
and focuses on the quantum mechanics of electrons in 
{\it nearly periodic potentials}.
The nearly periodic potentials arise either from 
uniform crystals in the presence of scalar
potentials varying slowly in space 
(for example, from impurities or gates),  
or from compositionally {\it graded}
crystals with or without applied potentials. 
In this Letter, I shall present a summary of new results on the
time-dependent properties of electrons in graded semiconductors.
Details of the calculations will be presented elsewhere.

{\it I. Semiclassical dynamics.}---I begin by discussing the 
semiclassical equations of motion for an electron in a graded
semiconductor.
Let $\epsilon_n({\bf k},c)$ denote the energy bands of a 
homogeneous alloy 
${\rm A_c B_{1-c}}$, which I assume are known. 
I shall consider a graded crystal 
with a composition $c({\bf r})$ varying slowly
on the scale of a characteristic lattice constant, in the 
presence of a slowly varying scalar 
potential $U({\bf r})$, and I shall for simplicity 
consider only noncomposite bands. 
The stationary states,
$ \psi_{n \alpha}({\bf r}) 
= \sum_{\bf l} \Phi^{\alpha}_{n {\bf l}}  
\ a_{n {\bf l}}({\bf r}-{\bf l}) , $
and energies, $\epsilon_{n \alpha}$,
of an electron in a graded crystal may be obtained by 
solving an eigenvalue problem,
\begin{equation}
\sum_{{\bf l}'} {\cal H}^n_{{\bf l l}'} \ \Phi_{n {\bf l}'}^\alpha
+ U({\bf l}) \ \Phi_{n{\bf l}}^\alpha
= \epsilon_{n \alpha} \ \Phi_{n{\bf l}}^\alpha ,
\label{eigenvalue problem}
\end{equation}
with an effective Hamiltonian
\begin{equation}
{\cal H}^n_{{\bf l l}'} \equiv {1 \over v} 
\int_{\rm \scriptscriptstyle BZ} d^3k \  \epsilon_n\big({\bf k},
c( {\scriptscriptstyle  {{\bf l}+{\bf l}' \over 2} }) \big)
e^{i {\bf k} \cdot ({\bf l}-{\bf l}')}  
\label{effective Hamiltonian}
\end{equation}
determined by the {\it local band structure} 
$\epsilon_n\big({\bf k},c({\bf r}) \big)$
of the graded crystal \cite{Geller and Kohn}.
Here $\alpha$ is an index labeling the different states 
associated with band $n$, 
and $v$ is the Brillouin zone volume.
The generalized Wannier functions $a_{n {\bf l}}$ 
in a nearly periodic
potential are labeled by a band index $n$   
and a lattice vector ${\bf l}$ about which the function 
is localized. They have the same
completeness, orthonormality, and band-diagonality 
properties as the standard 
Wannier functions for strictly periodic 
potentials \cite{GWF}. The use of generalized Wannier functions
makes it possible to construct the stationary 
states from a {\it single band},
and hence, I label them by the pair of indices $n$ and $\alpha$. 
This is possible as long as $c({\bf r})$ and 
$U({\bf r})$ are slowly varying.

The classical effective Hamiltonian for an electron
in band $n$ is 
$H = \epsilon_n({\bf k},c({\bf r})) + U({\bf r})$. 
The semiclassical equations of motion, regarding ${\bf r}$ and
$\hbar {\bf k}$ as conjugate variables, are
${\dot {\bf r}} = \hbar^{-1} [ \partial \epsilon_n({\bf k},c({\bf r}))
/ \partial {\bf k}] $
and
$\hbar {\dot {\bf k}} = - \nabla U({\bf r}) 
- [ \partial \epsilon_n({\bf k},c({\bf r}))
/ \partial c] \nabla c({\bf r})$.
The force includes a contribution 
from the composition gradient.
The effective Hamiltonian 
(\ref{effective Hamiltonian}) can be used to
show that these semiclassical equations also govern the quantum mechanical
motion of wave packets in a graded semiconductor.
This constitutes a generalization of Ehrenfest's theorem to the
motion of electrons in slowly graded crystals.

{\it II. Dynamic conductivity.}---The frequency-dependent conductivity
tensor for an electron in a state $\psi_{n \alpha}$ is given by
\end{multicols}
\begin{equation}
\sigma^{ij}(\omega) = {ie^2 \over m(\omega + is) V} \bigg[
 \delta^{ij}
- {1 \over m} \sum_{n' \alpha'}  
\bigg(
{ \langle \psi_{n \alpha} | p^i | \psi_{n' \alpha'} \rangle
\ \langle \psi_{n' \alpha'} | p^j | \psi_{n \alpha} \rangle
\over \epsilon_{n' \alpha'} - \epsilon_{n \alpha} 
- \hbar \omega - i \hbar s } 
+ { \langle \psi_{n \alpha} | p^i | \psi_{n' \alpha'} \rangle^*
\ \langle \psi_{n' \alpha'} | p^j | \psi_{n \alpha} \rangle^*
\over \epsilon_{n' \alpha'} - \epsilon_{n \alpha} 
+ \hbar \omega + i \hbar s} 
\bigg) \bigg],
\label{conductivity}
\end{equation}
where $m$ is the electron mass, $V$ is the volume of the crystal,
and $s$ is a positive infinitesimal.
The real part is 
\begin{equation}
\sigma_1^{ij}(\omega) = {\pi e^2 \hbar \over 2 m V }  
{\sum_{n' \alpha '}}'
f^{ij}_{n \alpha , n' \alpha'}     
\bigg[ 
\delta( \epsilon_{n' \alpha'} - \epsilon_{n \alpha} - \hbar \omega) 
+ \delta( \epsilon_{n' \alpha'} - \epsilon_{n \alpha} + \hbar \omega) 
\bigg]                      
+  { \pi e^2 \over m V} \bigg(  \delta^{ij}
- {\sum_{n' \alpha'}}'
f^{ij}_{n \alpha , n' \alpha'} \bigg) \ \delta (\omega) ,     
\label{real part}
\end{equation}
\begin{multicols}{2}
\noindent where 
\begin{equation}
f^{ij}_{n \alpha , n' \alpha'}
\ \equiv \ {2 \over m} { {\rm Re}
\  \langle \psi_{n \alpha} | p^i | \psi_{n' \alpha'} \rangle
\ \langle \psi_{n' \alpha'} | p^j | \psi_{n \alpha} \rangle 
\over  \epsilon_{n' \alpha'} - \epsilon_{n \alpha}  }
\label{oscillator strength}
\end{equation}
is the oscillator strength for an optical transition
between states
$\psi_{n \alpha}$
and
$\psi_{n' \alpha'}$.
The primed summation in (\ref{real part}) means that the 
$n' \alpha'  = n \alpha$
term is to be excluded.
Thus to calculate $\sigma^{ij}(\omega)$ we need, in addition to the
energies $\epsilon_{n \alpha}$ determined from
(\ref{eigenvalue problem}),  
also the momentum matrix
elements
$\langle \psi_{n \alpha} | {\bf p} | \psi_{n' \alpha'} \rangle
= \sum_{\bf l l'} (\Phi_{n {\bf l}}^\alpha)^*
\Phi_{n' {\bf l}'}^{\alpha'} \langle a_{n {\bf l}} |
{\bf p} | a_{n' {\bf l}'} \rangle$.
Because the generalized Wannier functions are localized, the
quantity
$\langle a_{n {\bf l}} |
{\bf p} | a_{n' {\bf l}'} \rangle$
depends only on the composition of the alloy near
${\bf l}$ and ${\bf l}'$. Hence, in a slowly graded
crystal,
\begin{eqnarray}
\langle \psi_{n \alpha} | {\bf p} | \psi_{n' \alpha'} \rangle
&=& {1 \over v} \sum_{\bf l l'} (\Phi_{n {\bf l}}^\alpha)^*
\Phi_{n' {\bf l}'}^{\alpha'} \nonumber \\
&\times &  \int d^3k  
\ e^{i {\bf k} \cdot ({\bf l}-{\bf l}')} 
{\bf p}_{n n'}\big( {\bf k},  
c({\scriptscriptstyle{{\bf l}+{\bf l}'\over 2}}) 
\big),
\end{eqnarray}
where
${\bf p}_{n n'}({\bf k},c) \equiv 
\langle \varphi_{n {\bf k}},c|{\bf p}|\varphi_{n' {\bf k}},c\rangle$
are the usual Bloch function matrix elements for a
homogeneous semiconductor with composition $c$.
Therefore, an evaluation of the dynamic conductivity does
{\it not} require a knowledge of the generalized Wannier functions,
which are difficult to calculate, but only of band
structures $\epsilon_n({\bf k},c)$ and momentum matrix
elements ${\bf p}_{n n'}({\bf k},c)$ of homogeneous alloys,
which are obtainable from conventional electronic structure\
calculations or experiments.  

{\it III. Sum rules.}---I shall consider  
{\it partial sums} of oscillator strengths
of the form
$ {\sum^\prime}_{\! \! \alpha'} f^{ij}_{n \alpha , n' \alpha'}$. 
When $n' \neq n$, this is the sum of 
interband oscillator strengths
between $\psi_{n \alpha}$ and all states $\psi_{n' \alpha'}$ 
in band $n'$, whereas, when
$n' = n$, it is the sum of all oscillator strengths 
between $\psi_{n \alpha}$ and all
{\it other} states in the same  band. 
In terms of the projection operator 
${\rm P}_{\! n} \equiv \sum_\alpha |\psi_{n \alpha} \rangle  
\langle \psi_{n \alpha} |$, I find
\begin{equation}
{\sum_{\alpha' }}' f^{ij}_{n \alpha, n' \alpha' }  
= { \big\langle \psi_{n \alpha} \big| x^i \ \! {\rm P}_{\! n'} \ \! p^j 
- p^j \ \! {\rm P}_{\! n'} \ \! x^i \big| \psi_{n \alpha} \big\rangle
\over i \hbar } .
\label{general sum rule}
\end{equation}
It is clear from (\ref{general sum rule})
that the partial sum depends only on the state 
$\psi_{n \alpha}$ and on the properties
of the crystal in the absence of applied fields. 
Note that
${\sum}'_{n' \alpha'} f^{ij}_{n \alpha, n' \alpha'} = \delta^{ij},$
as expected.

An {\it intraband sum rule} may be obtained 
from (\ref{general sum rule})
with $n' = n$,
\begin{equation}
\sum_{\alpha' \neq \alpha} f^{ij}_{n \alpha, n \alpha' }  
= { \big\langle \psi_{n \alpha} \big| [ x^i_n , p^j_n ] 
 \big| \psi_{n \alpha} \big\rangle
\over i \hbar },
\label{projected commutator}
\end{equation}
where
$O_n \equiv {\rm P}_{\! \! n} O {\rm P}_{\! \! n}$ 
denotes a projected operator.
This commutator can be
evaluated in a Wannier function basis, leading to
\begin{equation}
\sum_{\alpha' \ne \alpha} \ f^{ij}_{n \alpha , n \alpha'} 
= - {m \over \hbar^2}  
\ \sum_{{\bf l l}'} 
\big( \Phi_{n{\bf l}}^{\alpha} \big)^* 
{\cal H}^n_{\bf l l'}
\big( {\bf l}-{\bf l}' \big)^i
\big( {\bf l}-{\bf l}' \big)^j
\ \Phi_{n{\bf l}'}^{\alpha} .
\label{intraband sum rule}
\end{equation}
The sum of the oscillator strengths for transitions between
a given state 
$\psi_{n \alpha}$ and the other states in the same band 
is therefore proportional to the expectation value of the second moment
of (\ref{effective Hamiltonian}) in the state
$\Phi_{n {\bf l}}^\alpha$.

The result (\ref{intraband sum rule}) 
is valid for any state
$\psi_{n \alpha}$
in band $n$.
However, if the state 
is near in energy to a band edge, the sum rule 
simplifies to an expectation value of the local inverse
effective mass tensor, 
\begin{equation}
\sum_{\alpha' \ne \alpha} \ f^{ij}_{n \alpha , n \alpha'} 
= \ m \ \bigg( F_{n \alpha} \bigg|
\bigg( {1 \over m^*({\bf r}) }\bigg)^{ij}
\bigg| F_{n \alpha} \bigg) ,
\label{band edge intraband f sum rule}
\end{equation}
where
$\big(F \big| O \big| F \big) \equiv
\int d^3r \ F^* O F $
is the ordinary envelope function expectation value.

The intraband sum rule leads immediately to a sum rule for the
${\it intraband}$ contribution to $\sigma_1^{ij}(\omega)$, namely
\end{multicols}
\begin{equation}
\int_{\rm intra} d\omega \ \sigma_1^{ij}(\omega)  
=  - {\pi e^2 \over 2 V \hbar^2} 
\sum_{{\bf l l}'} 
\big( \Phi_{n{\bf l}}^{\alpha} \big)^* 
{\cal H}_{\bf l l'}
\big( {\bf l}-{\bf l}' \big)^i
\big( {\bf l}-{\bf l}' \big)^j
\ \Phi_{n{\bf l}'}^{\alpha} , 
\label{intraband conductivity sum rule}
\end{equation}
where the integration is to be taken from zero to 
the largest intraband frequency.

{\it Interband sum rules} may also be obtained 
from (\ref{general sum rule}) with $n' \neq n$.
Using a Wannier function basis, I find
\begin{equation}
\sum_{\alpha' } \ f^{ij}_{n \alpha , n' \alpha'} 
= \sum_{{\bf l l}'} 
\big( \Phi_{n{\bf l}}^{\alpha} \big)^* 
\ \Phi_{n{\bf l}'}^{\alpha}  
\ {1 \over v} \int_{\scriptscriptstyle \rm BZ} 
d^3k \ f_{nn'}^{ij}\big({\bf k},
c( {\scriptscriptstyle  {{\bf l}+{\bf l}' \over 2} }) \big)
\ e^{i {\bf k} \cdot ({\bf l}-{\bf l}')} ,
\label{interband f sum rule}
\end{equation}
\begin{multicols}{2}
\noindent where $f^{ij}_{nn'}({\bf k},c)$ 
is the conventional Bloch electron
oscillator strength.
For a state near in energy to a band edge,
the interband sum rule reduces to
\begin{equation}
\sum_{\alpha'} \ f^{ij}_{n \alpha , n' \alpha'} 
=  \big( F_{n \alpha} \big|
f^{ij}_{nn'} \big( 0 , c({\bf r}) \big)
\big| F_{n \alpha} \big) ,
\label{band edge interband f sum rule}
\end{equation}
an extremely simple form indeed.

{\it IV. Principle of spectroscopic stability for a graded 
semiconductor.}---I shall briefly 
discuss the physical origin of the new sum rules.
For this purpose I study the
stability of {\it double sums} of oscillator strengths of the form 
$S_{\! n n'}^{ij} \equiv \sum_\alpha \sum_{\alpha'} 
f^{ij}_{n \alpha, n' \alpha'}$
in the presence of perturbations.
Double sums of this type are commonplace in the theory 
of atomic spectra, where they
characterize the {\it total} optical absorption 
strength between two multiplets $n$ and $n'$, the
individual degenerate or nearly degenerate states in each 
multiplet being labeled by $\alpha$ and
$\alpha'$ respectively. The {\it invariance} of the total 
absorption strength between two
multiplets, under arbitrary unitary transformations among the 
degenerate or nearly degenerate states
in each multiplet, is known as the {\it principle of spectroscopic
stability} \cite{Condon and Shortley}. 
In the atomic physics context, these
unitary transformations usually arise from the 
application of a weak electric or magnetic field.
This principle is not immediately
applicable to our double sum, however, 
because $S_{\! n n'}^{ij}$ describes
the total absorption strength between {\it bands} of 
states which are not nearly 
degenerate. 

I have proved a stronger version 
of the stability principle by showing that
$S_{\! n n'}^{ij} = {\rm Tr}_n
( x^i {\rm P}_{\! n'}
p^j - p^j {\rm P}_{\! n'}  x^i) / i \hbar$ ,
where the partial trace acts in the subspace 
spanned by band $n$.
The invariance of
$S_{\! n n'}^{ij}$
under arbitrary unitary 
transformations within the band $n$ and
independently within the band $n'$ is now evident. 
In particular, 
$S_{\! n n'}^{ij}$
will be invariant
under the action of slowly varying perturbations. 
Because 
$S_{\! n n'}^{ij}$
is conserved in going from a periodic potential
to a nearly periodic one, the optical spectrum of the latter is
related to that of the former by an {\it intraband redistribution} of
the transition strengths.
This fact implies the existence of general sum 
rules for the oscillator strengths
of an electron in a graded semiconductor, 
as demonstrated above.

{\it V. Width of the Drude peak.}---The intraband optical absorption
spectrum of an electron of momentum ${\bf k}$ in a uniform crystal
with composition $c$ is
\begin{equation}
\sigma_1^{ij}(\omega ) = {\pi e^2 \over \hbar^2 V}
{\partial^2 \epsilon_n({\bf k},c) \over \partial k_i \partial k_j}
\delta(\omega ).
\end{equation}
In the presence of an applied potential $U({\bf r})$ or a
composition gradient or both, the width of this Drude peak is
broadened and its integrated strength 
(\ref{intraband conductivity sum rule}) changes.
Here I shall present a general expression for the width
$ \delta \omega$, defined by
\begin{equation}
{\delta \omega}^2 \equiv { \int d\omega \ \omega^2 \sigma_1^{ij} 
\over \int d\omega \ \sigma_1^{ij} },
\label{width definition}
\end{equation}
where $\sigma_1^{ij}$ is the real part of the conductivity for a
state $\psi_{n \alpha}$, and the integrations are to include intraband
contributions only. The denominator of (\ref{width definition})
is determined by
the intraband sum rule 
(\ref{intraband conductivity sum rule}), whereas for
the numerator I find
\end{multicols}
\begin{equation}
\int d\omega \ \omega^2 \sigma_1^{ij} 
= - {\pi e^2 \over m^2 \hbar^2 V}
\sum_{\bf l l'} (\Phi_{n {\bf l}}^\alpha)^*
\ [[ H,P^i],P^j]_{\bf l l'} \ \Phi_{n {\bf l}'}^\alpha,
\label{width}
\end{equation}
where 
$H_{\bf l l'} \equiv {\cal H}_{\bf l l'} + 
U({\bf l}) \delta_{\bf l l'}$ and
$P^i_{\bf l l'} \equiv (m / \hbar v)
\int d^3k \ [\partial \epsilon_n({\bf k}, 
c({\scriptscriptstyle{{\bf l}+{\bf l}' \over 2}}) 
) / \partial k_i ] e^{i {\bf k} \cdot ({\bf l}-{\bf l}')}.$
The width depends only on the state $\Phi_{n {\bf l}}^\alpha$
and can be computed directly from (\ref{width}) once
$\Phi_{n {\bf l}}^{\alpha}$ is known. 
In the effective mass regime (\ref{width}) simplifies to
\begin{equation}
\int d\omega \ \omega^2 \sigma_1^{ij} 
= {\pi e^2 \over V} 
\bigg( F_{n \alpha} \bigg|
\bigg({1 \over m^*({\bf r})}\bigg)^{ik}
\bigg({1 \over m^*({\bf r})}\bigg)^{jl}
\nabla_k \nabla_l 
\big[ {\cal E}({\bf r}) + U({\bf r}) \big] \bigg| F_{n \alpha} \bigg),
\end{equation}
\begin{multicols}{2}
\noindent where ${\cal E}({\bf r}) 
\equiv \epsilon_n(0,c({\bf r}))$
is the energy of the local band minimum.

{\it VI. Low-frequency dynamics.}---Consider an electron in
a state of energy $\epsilon_{n \alpha}$ near a band edge.
At frequencies much less than the characteristic energy
gap $\hbar {\bar \omega}$ to the nearest state in the same 
band with appreciable oscillator strength, the conductivity
is purely imaginary and is given by
\begin{equation}
\sigma^{ij}(\omega) = - {i e^2 \omega \over V 
{\bar \omega}^2 }
\bigg( F_{n \alpha} \bigg|
\bigg({1 \over m^*({\bf r})}\bigg)^{ij}
\bigg| F_{n \alpha} \bigg).
\label{low frequency conductivity}
\end{equation}
Typically, $\hbar {\bar \omega}$ will be equal to the level
spacing near $\epsilon_{n \alpha}$.

The low-frequency conductivity 
(\ref{low frequency conductivity}) can also be obtained 
from the semiclassical equation of motion
$m^* (d^2z/dt^2) + {\textstyle {1 \over 2}} (dm^*/dz)
(dz/dt)^2
+ d{\cal E}/dz = e E(t)$,
where $m^*(z)$ is the position-dependent effective mass and
$E(t) = E_0 e^{-i \omega t}$ is the applied electric field.
To find the response of the electron to the driving field I 
write $z(t) = z_0(t) + z_1(t)$, where $z_0$ is the semiclassical
trajectory in the absence of the driving field with initial
conditions corresponding to a solution with energy $\epsilon$.
Expanding about the unperturbed motion and using the fact that
$m^*(z)$ is slowly varying leads to
\begin{equation}
m^*(z_0) {d^2 z_1 \over dt^2} + {d m^*(z_0) \over dz} 
{d z_0 \over dt} {d z_1 \over dt}
+ {\cal E}''(z_0) z_1 = e E(t).
\label{linearlized equation of motion}
\end{equation}
Note that the coefficients of this linearized equation of motion
are time dependent.
The low-frequency response is determined by the solution of
(\ref{linearlized equation of motion}) with the driving
frequency $\omega$ much smaller than the frequency of the
unperturbed motion at energy $\epsilon$. This means that
the electron completes many closed orbits during a single
driving cycle. Because the motion is generally anharmonic,
the response of the electron will contain a low frequency
component at $\omega$ as well as higher harmonics. I shall
determine the low-frequency component by averaging the
rapidly varying coefficients over a period $T$ of the
unperturbed motion. Thus,
\begin{eqnarray}
&& \big\langle m^*(z_0) \big\rangle {d^2 z_1 \over dt^2} 
+ \bigg\langle {d m^*(z_0) \over dz} 
{d z_0 \over dt} \bigg\rangle {d z_1 \over dt}
+ \big\langle {\cal E}''(z_0) \big\rangle z_1 \nonumber \\
&& = e E(t),
\label{averaged equation of motion}
\end{eqnarray}
where $\langle f(t) \rangle \equiv T^{-1} 
\int_0^T dt \ f(t)$.
Therefore, the low-frequency response is that of a
damped harmonic oscillator,
$ {\ddot x} + {\dot x} / \tau + \Omega^2 x = e E(t)$,
where
$\tau \equiv \langle m^* \rangle / \langle (dm^*/dz) {\dot z_0}
\rangle $
and 
$\Omega^2 \equiv \langle{\cal E}'' \rangle 
/ \langle m^* \rangle$.
Both $\tau$ and $\Omega$ depend on the energy $\epsilon$
of the unperturbed orbit.
Assuming $\omega \ll \Omega$ and 
$\omega \ll \tau^{-1}$,
the low-frequency conductivity at this energy is
\begin{equation}
\sigma^{zz}(\omega) = - {i e^2 \omega \over
V \Omega^2 \langle m^* \rangle },
\end{equation}
in close
correspondence with (\ref{low frequency conductivity}).

{\it VII. Absorption spectrum in a quantum well.}---As an 
application of my results
I shall calculate the optical absorption spectrum
of an electron in a slowly graded
${\rm Al_c Ga_{1-c} As}$
parabolic quantum well
\cite{Gossard etal}.
In particular, we shall examine the effects of the position-dependent
band structure on the intraband oscillator strengths and selection 
rules. The eigenstates of interest here are near in energy
to the minimum of the local conduction band and hence
may be described by the effective mass 
Hamiltonian \cite{Bastard etal,Geller and Kohn}
$ H = - {\textstyle {1 \over 2}}\hbar^2 \nabla_i
[1 / m^*({\bf r}) ]^{ij} \nabla_j
+ {\cal E}({\bf r}). $ 
Here ${\cal E}({\bf r}) = {\textstyle {1 \over 2}} m^* \omega_0^2 z^2,$ 
where $m^*$
is the electron effective mass in GaAs, and 
$\hbar \omega_0 $
is the energy level spacing at the bottom of the well.
The laser field is assumed to be polarized in the $z$ direction.
The effective mass for an electron at the $\Gamma$ point of
${\rm Al_c Ga_{1-c} As}$
is known to be well-described by a linear interpolation between
the effective mass of
GaAs and  the  
$\Gamma$-point effective mass
of AlAs.
The position-dependent
effective mass in the quantum well may be written as 
$ m^*(z) = m^* 
[ 1 + \eta (z / \ell)^2 ],$
where $\ell^2 \equiv \hbar / m^* \omega_0$,
and where
$\eta $
is a dimensionless quantity characterizing
the relative change in the effective mass over a length
$\ell$.
In an 
${\rm Al_c Ga_{1-c} As}$
parabolic quantum well with $\hbar \omega_0 \approx 1 \ {\rm meV}$,
it can be shown that
$\eta \ll 1$.
Thus, the effective-mass gradient 
may be treated perturbatively.
To this end, I write 
$ H = H^0 + H^1$,
where
$ H^1 \equiv \eta \hbar \omega_0 ( {\textstyle{1 \over 2}}  z^2
\partial^2_z  
+ z \partial_z ).$
The energies and normalized envelope functions of
$H^0$ 
are given by
$ \epsilon_{j{\bf k}}^0 = ( j + {\textstyle{1 \over 2}} ) \hbar \omega_0
+ \hbar^2 {\bf k}^2 / 2 m^* $
and
$F_{j {\bf k}}^0({\bf r}) = (2^j j! \pi^{1 \over 2} \ell
L^2 )^{-{1 \over 2}}  e^{i {\bf k} \cdot {\bf r}} 
e^{-{1 \over 2}(z/\ell)^2}
 H_j(z / \ell ),$ 
where ${\bf k}$
is a wavevector in the plane of the quantum well, the
$H_j$ are Hermite polynomials,
and where I have used periodic boundary conditions
in the $x$ and $y$ directions.
The envelope functions
determine the actual eigenfunctions of the electron in the $n$th
band through 
$ \psi_{n j {\bf k}}({\bf r}) = \sum_{\bf l} F_{j{\bf k}}({\bf l})
\ a_{n{\bf l}}({\bf r}-{\bf l}) $.

I shall calculate the $zz$ components of the
intraband oscillator strengths, 
$ f_{j{\bf k},j'{\bf k}'} = 
2m ( \epsilon_{j'{\bf k}'} - \epsilon_{j{\bf k}})
| \langle \psi_{nj{\bf k}} | z 
| \psi_{nj'{\bf k}'} \rangle 
|^2 / \hbar^2 ,$ 
to first order in $\eta$.
To this order, the perturbed eigenvalues are
$ \epsilon_{j{\bf k}} = \epsilon_{j{\bf k}}^0 - {1 \over 4} \eta
( j^2 + j + {3 \over 2} ) \hbar \omega_0 $.
The nonvanishing intraband oscillator strengths to order $\eta$ are
$f_{j {\bf k}; j+1, {\bf k}}  = 
(m / m^*) [ j + 1 - \eta ( {\textstyle{1 \over 2}} j^2
+ j + {\textstyle{1 \over 2}} )]$ 
and
$ f_{j {\bf k}; j-1, {\bf k}} = 
(m /m^*) [ - j   + \eta ( {\textstyle{1 \over 2}} j^2 )].$
The selection rules $j \rightarrow j \pm 1$
are therefore unchanged to first order in the effective mass gradient.
However, the oscillator strengths are indeed modified, and
the optical absorption frequencies,
$ \epsilon_{j+1,{\bf k}} - \epsilon_{j,{\bf k}} 
= [ 1 - \eta {\textstyle{1 \over 2}} 
(j+1)] \hbar \omega_0 $,
are decreased.
The sum of the intraband oscillator strengths is 
$(m /m^*) [ 1 - \eta
(j + {\textstyle{1 \over 2}} )],$
which is 
in agreement with (\ref{band edge intraband f sum rule})
to order $\eta$.

It is known that deviations from perfect parabolic confinement
and the existence of a position-dependent effective mass both modify the
optical absorption spectrum of an ideal parabolic quantum
well, by changing the level spacing and oscillator strengths.
Is it possible to separate the effects of these two perturbations,
which are always present in real quantum wells?
The above analysis shows that the {\it integrated} intraband
absorption strength depends on the presence of the 
position-dependent effective mass only, and is therefore a direct 
probe of this subtle band-structure effect.

It is a pleasure to thank Walter Kohn for useful discussions on this subject.

\end{multicols}


\begin{references}

\bibitem{Bastard etal} See, for example, G. Bastard, 
J. A. Brum, and R. Ferreira, in 
{\it Solid State Physics: Advances in Research and Applications}, edited by
H. Ehrenreich and D. Turnbull (Academic Press, New York, 1991), Vol. 44.

\bibitem{Geller and Kohn} M. R. Geller and W. Kohn, 
Phys. Rev. Lett. {\bf 70}, 3103 (1993).

\bibitem{Mahan} For a good discussion see G. D. Mahan,
{\it Many-Particle Physics} (Plenum Press, New York, 1990).

\bibitem{GWF} M. R. Geller and W. Kohn, Phys. Rev. B {\bf 48}, 14085 (1993).

\bibitem{Condon and Shortley} E. U. Condon and G. H. 
Shortley, {\it The Theory of Atomic Spectra}
(Cambridge University Press, New York, 1967).

\bibitem{Gossard etal} For a review, see 
A. C. Gossard, M. Sundaram, and P. F. Hopkins,
in {\it Semiconductors and Semimetals}, edited by  Willardson and 
Beer (Academic Press, New York, 1993), Vol. 40.

\end{references}
\end{document}